\documentclass[12pt]{article}

\usepackage{amsmath,amssymb}

\begin{document}


\begin{center}

{\Large \bf Series expansion in fractional calculus and fractional
differential equations}\vskip 7mm

{\large Ming-Fan Li$^{\ast}$, Ji-Rong Ren$^{\ddag}$, Tao
Zhu$^{\S}$}\vskip 3mm

{\it Institute of Theoretical Physics,\\
Lanzhou University, Lanzhou, 730000, China}\\

$^{\ast}${E-mail: limf07@lzu.cn}\\
$~~$$^{\ddag}${E-mail: renjr@lzu.edu.cn}\\
$^{\S}${E-mail: zhut05@lzu.cn}

\end{center}\vskip 3mm

\begin{abstract}
Fractional calculus is the calculus of differentiation and
integration of non-integer orders. In a recent paper (Annals of
Physics {\bf 323} (2008) 2756-2778), the Fundamental Theorem of
Fractional Calculus is highlighted. Based on this theorem, in this
paper we introduce fractional series expansion method to fractional
calculus. We define a kind of fractional Taylor series of infinitely
fractionally-differentiable functions. By using this kind of
fractional Taylor series, we give a fractional generalization of
hypergeometric functions and derive corresponding differential
equations. For finitely fractionally-differentiable functions, we
observe that the non-infinitely fractionally-differentiability is
due to more than one fractional indices. We expand functions with
two fractional indices and illustrate how this kind of series
expansion can help to solve fractional differential equations.

\end{abstract}\vskip 3mm


Fundamental Theorem of Fractional Calculus;

Series expansion;

Fractional hypergeometric function;

Fractional differential equation.

\vskip 2mm






\section{Introduction}

Fractional calculus is the calculus of differentiation and
integration of non-integer orders. During last three decades or so,
fractional calculus has gained much attention due to its
demonstrated applications in various fields of science and
engineering
\cite{oldham,samko,miller,podlubny,kilbs,carpinteri,hilfer,west,zaslav,metzler,hermann,agrawal,entropy,varCal,physScr,jmp,iomin,FundaTheorFC}.
There are many good textbooks of fractional calculus and fractional
differential equations, such as
\cite{oldham,samko,miller,podlubny,kilbs}. For various applications
of fractional calculus in physics, see
\cite{carpinteri,hilfer,west,zaslav,metzler,hermann,agrawal,entropy,varCal,physScr,jmp,iomin,FundaTheorFC}
and references therein.

In Calculus, the Fundamental Theorem of Calculus (Newton-Leibniz
Theorem) is of fundamental importance, matching its name. For
fractional calculus, an analogous theorem has been highlighted
recently in a paper \cite{FundaTheorFC}. When trying to construct a
consistent fractional vector calculus, V. E. Tarasov observed that
many of fundamental problems can be solved by using a generalization
of the Fundamental Theorem of Calculus.

Series expansion is an important tool to calculus. Particularly,
series expansion plays an important role in solving some
differential equations, such as the hypergeometric differential
equations \cite{wzx,wiki_HyperF,wiki_ConfHyperF}. However,
fractional series expansion has not yet been introduced to
fractional calculus.

In this paper, based on the Fundamental Theorem of Fractional
Calculus, we will introduce fractional series expansion method to
fractional calculus. We will define a kind of fractional Taylor
series of infinitely fractionally-differentiable functions. By using
of this kind of fractional Taylor series, we will give a fractional
generalization of hypergeometric functions and derive corresponding
differential equations. For finitely fractionally-differentiable
functions, we observe that the non-infinitely
fractionally-differentiability is due to more than one fractional
indices. We will expand functions with two fractional indices and
illustrate how this kind of series expansion can help to solve
fractional differential equations.

The structure of this paper is as follows. In Section \ref{FC}, we
briefly review fractional derivative, fractional integral and the
Fundamental Theorem of Fractional Calculus. In Section \ref{FTS}, we
introduce the fractional Taylor series of an infinitely
fractionally-differentiable function and give some examples. In
Section \ref{FHyperF}, we make a generalization of the
hypergeometric functions. In Section \ref{FtwoIn}, we discuss
finitely fractionally-differentiable functions. In Section
\ref{Summary}, we give our summary.

\section{Fractional Caculus}\label{FC}

In this section, we briefly review the definitions of fractional
integral and fractional derivative, and the Fundamental Theorem of
Fractional Calculus. For more details, see
\cite{oldham,samko,miller,podlubny,kilbs}.

\subsection{Fractional integral and fractional derivative}

There are many ways to define fractional derivative and fractional
integral. Most of them are based on the idea that we can generalize
the equation
\begin{equation}\label{dpower}
    \frac{d^n x^m}{d x^n}=\frac{m!}{(m-n)!}x^{m-n},~~~~n\in N,
\end{equation}
by replacing each factorial by a gamma function, to
\begin{equation}\label{}
    \frac{d^{\alpha} x^{\beta}}{d
    x^{\alpha}}=\frac{\Gamma(\beta+1)}{\Gamma(\beta-\alpha+1)}x^{\beta-\alpha},~~~~\alpha>0.
\end{equation}
In terms of integral operation $I$, the idea is to generalize
\begin{equation}\label{ipower}
    I^n x^m=\frac{m!}{(m+n)!}x^{m+n}
\end{equation}
to
\begin{equation}\label{}
    I^{\alpha}
    x^{\beta}=\frac{\Gamma(\beta+1)}{\Gamma(\alpha+\beta+1)}x^{\beta+\alpha}.
\end{equation}

The most commonly-used fractional integral is the Riemann-Liouville
fractional integral (RLFI).

Let $f(x)$ be a function defined on the interval $[a,b]$. Let
$\alpha$ be a positive real.

The right Riemann-Liouville fractional integral (right-RLFI) is
defined by
\begin{equation}\label{}
    I^{\alpha}_{a|x}f(x)=\frac{1}{\Gamma(\alpha)}\int_a^x
    (x-\xi)^{\alpha-1}f(\xi)d\xi,
\end{equation}
and the left Riemann-Liouville fractional integral (left-RLFI) is
defined by
\begin{equation}\label{}
    I^{\alpha}_{x|b}f(x)=\frac{1}{\Gamma(\alpha)}\int_x^b
    (\xi-x)^{\alpha-1}f(\xi)d\xi.
\end{equation}

We note that in this paper our use of notations ``right" and ``left"
is different from the common use, for reasons that will be clear
later.

For fractional derivatives, the Caputo fractional derivative (CFD)
is a commonly-used one. Let $n\equiv [\alpha]+1$.

The right-CFD and left-CFD are defined, respectively by
\begin{equation}\label{}
    D^{C,\alpha}_{a|x}f(x)=I^{n-\alpha}_{a|x}\frac{d^n}{dx^n}f(x)=\frac{1}{\Gamma(n-\alpha)}\int_a^x
    (x-\xi)^{n-\alpha-1}\frac{d^n}{d\xi^n}f(\xi)d\xi,
\end{equation}
\begin{equation}\label{}
    D^{C,\alpha}_{x|b}f(x)=I^{n-\alpha}_{x|b}(-\frac{d}{dx})^nf(x)=\frac{1}{\Gamma(n-\alpha)}\int_x^b
    (\xi-x)^{n-\alpha-1}(-\frac{d}{d\xi})^n f(\xi)d\xi.
\end{equation}

One can check that the above definitions really generalize
(\ref{dpower}) and (\ref{ipower}), and give
\begin{eqnarray}\label{relation1}
  D^{C,\alpha}_{a|x}(x-a)^{\beta} &=& \frac{\Gamma(\beta+1)}{\Gamma(\beta-\alpha+1)}(x-a)^{\beta-\alpha},~~~\beta\neq 0,1,...,[\alpha];  \\
  D^{C,\alpha}_{x|b}(b-x)^{\beta} &=& \frac{\Gamma(\beta+1)}{\Gamma(\beta-\alpha+1)}(b-x)^{\beta-\alpha},~~~\beta\neq 0,1,...,[\alpha]; \\
  I^{\alpha}_{a|x}(x-a)^{\beta} &=& \frac{\Gamma(\beta+1)}{\Gamma(\beta+\alpha+1)}(x-a)^{\beta+\alpha};  \\
  I^{\alpha}_{x|b}(b-x)^{\beta} &=& \frac{\Gamma(\beta+1)}{\Gamma(\beta+\alpha+1)}(b-x)^{\beta+\alpha}.
\end{eqnarray}
Especially, the Caputo fractional derivative on a constant
($\beta=0$) yields zero,
\begin{equation}\label{}
    D^{C,\alpha}_{a|x}\cdot 1=0,
\end{equation}
\begin{equation}\label{relation6}
    D^{C,\alpha}_{x|b}\cdot 1=0.
\end{equation}
This simple property is decisive in the fractional series expansion
and in our preference of the Caputo fractional derivative to another
commonly-used fractional derivative, the Riemann-Liouville
fractional derivative, whose operation on a constant gives not zero,
\begin{equation}\label{}
    D^{RL,\alpha}_{a|x}\cdot
    1=\frac{(x-a)^{-\alpha}}{\Gamma(1-\alpha)}, ~~~0<\alpha<1,
\end{equation}
\begin{equation}\label{}
    D^{RL,\alpha}_{x|b}\cdot 1=\frac{(b-x)^{-\alpha}}{\Gamma(1-\alpha)}, ~~~0<\alpha<1.
\end{equation}

\subsection{Fundamental Theorem of Fractional Calculus}

The Fundamental Theorem of Calculus (Newton-Leibniz Theorem) is
\begin{equation}\label{FT1}
    \int_a^b dx \frac{d}{dx}f(x)=f(b)-f(a),
\end{equation}
\begin{equation}\label{FT2}
    \frac{d}{dx} \int_a^x f(\xi) d\xi=f(x).
\end{equation}
This theorem means that the derivative operation is inverse to the
integral operation, and vice versa.

In fractional calculus, an analogous theorem exists
\cite{kilbs,FundaTheorFC}.

\noindent\emph{Fundamental Theorem of Fractional Calculus}.

\noindent(1) Let $\alpha>0$ and let $f(x)\in L_{\infty}(a,b)$ or
$f(x)\in C[a,b]$. Then
\begin{eqnarray}
  D^{C,\alpha}_{a|x}I^{\alpha}_{a|x}f(x) &=& f(x), \\
  D^{C,\alpha}_{x|b}I^{\alpha}_{x|b}f(x) &=& f(x).
\end{eqnarray}

\noindent(2) Let $0<\alpha<1$. If $f(x)\in AC[a,b]$ or $f(x)\in
C[a,b]$. Then
\begin{eqnarray}
  I^{\alpha}_{a|x}D^{C,\alpha}_{a|x}f(x) &=& f(x)-f(a), \\
  I^{\alpha}_{x|b}D^{C,\alpha}_{x|b}f(x) &=& f(x)-f(b).
\end{eqnarray}

Here $L_{\infty}(a,b)$ is the set of those Lebesgue measurable
functions $f$ on $(a,b)$ for which $\|f\|_{\infty}<{\infty}$, where
$\|f\|_{\infty}=$ ess sup$_{a\leq x\leq b}|f(x)|$. Here ess
sup$|f(x)|$ is the essential maximum of the function $|f(x)|$.
$AC[a,b]$ is the space of functions $f$ which are absolutely
continuous on $[a,b]$. $C[a,b]$ is the space of continuous functions
$f$ on $[a,b]$ with the norm $\|f\|_C=\max_{x\in[a,b]}|f(x)|$.

The proof of this theorem can be obtained from \cite{kilbs}, in
which these results are included in Lemma 2.21 and Lemma 2.22 there.

So, by this theorem, one can say that the right (left) Caputo
fractional derivative operation and the right (left)
Riemann-Liouville fractional integral operation are inverse to each
other.

\section{Fractional Taylor series of infinitely fractionally-differentiable functions}\label{FTS}

In this section, we will introduce fractional series expansion
method to fractional calculus and define a kind of fractional Taylor
series.

We observe that
\begin{equation}\label{}
    f(x)=f(a)+D^{\alpha}_{a|x}f(x)\big|_{x=\xi}\cdot (I^{\alpha}_{a|x}\cdot
    1),
\end{equation}
where $a<\xi<x$, and $\xi$ varies with the integral upper bound.
Here and after, $D^{\alpha}_{a|x}$ denotes the right Caputo
fractional derivative $D^{C,\alpha}_{a|x}$.

The corresponding formula in integer-order calculus is
\begin{equation}\label{}
    f(x)=f(a)+\frac{df}{dx}\bigg|_{x=\xi}\cdot(x-a),
\end{equation}
which is the Lagrange Mean Value Theorem.

Make a step further,
\begin{equation}\label{}
    f(x)=f(a)+D^{\alpha}_{a|x}f(x)\big|_{x=a}\cdot (I^{\alpha}_{a|x}\cdot
    1)+D^{\alpha}_{a|x}D^{\alpha}_{a|x}f(x)\big|_{x=\xi}\cdot (I^{\alpha}_{a|x}I^{\alpha}_{a|x}\cdot
    1).
\end{equation}

The corresponding formula in integer-order calculus is
\begin{equation}\label{}
    f(x)=f(a)+\frac{df}{dx}\bigg|_{x=a}\cdot(x-a)+(\frac{d}{dx})^2
    f\bigg|_{x=\xi}\cdot\frac{1}{2}(x-a)^2.
\end{equation}

And so on. One can extend this expansion to infinite order if the
function is sufficiently smooth.

Based on this observation a definition of a formal fractional Taylor
series expansion can be made.

\noindent{\bf Definition 1.a.} Let $f(x)$ be a function defined on
the right neighborhood of $a$, and be an infinitely
fractionally-differentiable function at $a$, that is to say, all
$(D^{\alpha}_{a|x})^m f(x)$($m=0,1,2,3..$) exist, and are not
singular at $a$. The formal fractional right-RL Taylor series of a
function is
\begin{equation}\label{}
    f(x)=\sum_{m=0}^{\infty} (D^{\alpha}_{a|x})^m f(x)\big|_{x=a}\cdot [(I^{\alpha}_{a|x})^m\cdot 1].
\end{equation}
Explicitly,
\begin{equation}\label{}
    (I^{\alpha}_{a|x})^m\cdot 1=\frac{1}{\Gamma(m\alpha+1)}(x-a)^{m\alpha}.
\end{equation}

\noindent{\bf Definition 1.b.} Let $f(x)$ be a function defined on
the left neighborhood of $b$, and be an infinitely
fractionally-differentiable function at $b$, that is to say, all
$(D^{\alpha}_{x|b})^m f(x)$($m=0,1,2,3..$) exist, and are not
singular at $b$. The formal fractional left-RL Taylor series of a
function is
\begin{equation}\label{}
    f(x)=\sum_{m=0}^{\infty} (D^{\alpha}_{x|b})^m f(x)\big|_{x=b}\cdot [(I^{\alpha}_{x|b})^m\cdot 1].
\end{equation}
Explicitly,
\begin{equation}\label{}
    (I^{\alpha}_{x|b})^m\cdot 1=\frac{1}{\Gamma(m\alpha+1)}(b-x)^{m\alpha}.
\end{equation}

In the above definitions, $D^{\alpha}_{a|x}$ is the right Caputo
fractional derivative $D^{C,\alpha}_{a|x}$; $D^{\alpha}_{x|b}$ is
the left Caputo fractional derivative $D^{C,\alpha}_{x|b}$.
$I^{\alpha}_{a|x}$ and $I^{\alpha}_{x|b}$ are right- and left-
Riemann-Liouvelle fractional integral, respectively.

\noindent{\bf Remark 1.} One can easily check the formal correctness
of the expansions by using of the Fundamental Theorem of Fractional
Calculus, or the relations (\ref{relation1})-(\ref{relation6}). For
rigorous validity, convergence is required.

\noindent{\bf Remark 2.} Series expansion has played an important
role in Calculus, particularly in solving differential equations.
However, fractional series expansion has not yet been introduced to
fractional calculus. This is because a pre-requisite that makes
fractional series expansion possible is the Fundamental Theorem of
Fractional Calculus, which is only recently proved and highlighted
\cite{kilbs,FundaTheorFC}.

\noindent{\bf Remark 3.} We may expect that fractional series
expansion will shed new light on fractional calculus, especially the
field of fractional differential equations. In the next section, we
will use this expansion to define a fractional generalization of
hypergeometric functions and discuss their differential equations.

\noindent{\bf Remark 4.} The fractional Taylor series is defined for
infinitely fractionally-differentiable functions. Finitely
fractionally-differentiable functions will be discussed in Section
\ref{FtwoIn}.

In the following, we give some simple examples of fractional Taylor
series.

\noindent{\bf Example 1.} $(x-a)^{\beta}$, with no $l\in N$
satisfying $\beta=l\alpha$.
\begin{equation}\label{}
    (D^{\alpha}_{a|x})^m(x-a)^{\beta}=\frac{\Gamma(\beta+1)}{\Gamma(\beta-m\alpha+1)}(x-a)^{\beta-m\alpha},
\end{equation}
For large $m$, the derivative will be singular at $a$. So we cannot
make the expansion.

\noindent{\bf Example 2.}
\begin{equation}\label{}
    e^{(x-a)^{\alpha}}=\sum_{m=0}^{\infty}\frac{1}{m!}(x-a)^{m\alpha}=\sum_{m=0}^{\infty}\frac{\Gamma(m\alpha+1)}{m!}\frac{1}{\Gamma(m\alpha+1)}(x-a)^{m\alpha}.
\end{equation}

\noindent{\bf Example 3.} The Mittag-Leffler function
$E_{\alpha}((x-a)^{\alpha})$, which satisfies
\begin{equation}\label{}
    D^{\alpha}_{a|x}E_{\alpha}((x-a)^{\alpha})=E_{\alpha}((x-a)^{\alpha}),~~~~~~~~~~~E_{\alpha}(0)=1,
\end{equation}
It is the fractional analogue of $exp(x)$. For arbitrary $m$,
$(D^{\alpha}_{a|x})^m E_{\alpha}((x-a)^{\alpha})|_{x=a}=1$. So,
\begin{equation}\label{}
    E_{\alpha}((x-a)^{\alpha})=\sum_{m=0}^{\infty}[(I^{\alpha}_{a|x})^m\cdot
1]=\sum_{m=0}^{\infty}\frac{1}{\Gamma(m\alpha+1)}(x-a)^{m\alpha}.
\end{equation}
Notice the difference between $e^{(x-a)^{\alpha}}$ and
$E_{\alpha}((x-a)^{\alpha})$.

\noindent{\bf Example 4.} $cos_{\alpha}((x-a)^{\alpha})$ and
$sin_{\alpha}((x-a)^{\alpha})$.
\begin{equation}\label{}
    cos_{\alpha}((x-a)^{\alpha})=1-\frac{(x-a)^{\alpha\cdot 2}}{\Gamma(\alpha\cdot
    2+1)}+\frac{(x-a)^{\alpha\cdot 4}}{\Gamma(\alpha\cdot 4+1)}-\frac{(x-a)^{\alpha\cdot 6}}{\Gamma(\alpha\cdot
    6+1)}+...
\end{equation}
\begin{equation}\label{}
    sin_{\alpha}((x-a)^{\alpha})=\frac{(x-a)^{\alpha}}{\Gamma(\alpha+1)}-\frac{(x-a)^{\alpha\cdot 3}}{\Gamma(\alpha\cdot 3+1)}+\frac{(x-a)^{\alpha\cdot 5}}{\Gamma(\alpha\cdot
    5+1)}-...
\end{equation}
They satisfy
\begin{eqnarray}
  D^{\alpha}_{a|x}sin_{\alpha}((x-a)^{\alpha}) &=&
  cos_{\alpha}((x-a)^{\alpha}), \\
  D^{\alpha}_{a|x}cos_{\alpha}((x-a)^{\alpha}) &=&
  -sin_{\alpha}((x-a)^{\alpha}).
\end{eqnarray}

\section{Fractional hypergeometric function}\label{FHyperF}

In the section, we will define a fractional generalization of the
hypergeometric functions.

Let us first consider a fractional generalization of the confluent
hypergeometric differential equation:
\begin{equation}\label{}
    z^{\alpha}(D^{\alpha}_{0|z})^2
    y+(c-z^{\alpha})D^{\alpha}_{0|z}y-ay=0.
\end{equation}
Here $a$ and $c$ are complex parameters. When $\alpha=1$, this is
the ordinary confluent hypergeometric equation.

Introducing the fractional Taylor series
\begin{equation}\label{yFTS}
    y(z)=\sum_{k=0}^{\infty}c_k z^{\alpha\cdot k},
\end{equation}
and substituting, we get the ratio of successive coefficients
\begin{eqnarray}
  \frac{c_{k+1}\cdot\Gamma(k\alpha+\alpha+1)}{c_{k}\cdot\Gamma(k\alpha+1)} &=& \frac{a+\frac{\Gamma(k\alpha+1)}{\Gamma(k\alpha-\alpha+1)}}{c+\frac{\Gamma(k\alpha+1)}{\Gamma(k\alpha-\alpha+1)}},\\
  \frac{c_1\cdot\Gamma(\alpha+1)}{c_0} &=& \frac{a}{c}.
\end{eqnarray}
Thus we get a solution of the above differential equation,
\begin{equation}\label{fconfhyperf}
    y(z)=\sum_{k=0}^{\infty}\frac{(a)^{\alpha}_k}{(c)^{\alpha}_k}\frac{1}{\Gamma(k\alpha+1)}z^{\alpha\cdot k}.
\end{equation}
Here $(a)^{\alpha}_k$ is defined as
\begin{eqnarray}
  (a)^{\alpha}_0 &=& 1, ~~~~~(a)^{\alpha}_1=a, \nonumber\\
  (a)^{\alpha}_k &=&
  \bigg(a+\frac{\Gamma(k\alpha-\alpha+1)}{\Gamma(k\alpha-2\alpha+1)}\bigg)\bigg(a+\frac{\Gamma(k\alpha-2\alpha+1)}{\Gamma(k\alpha-3\alpha+1)}\bigg)...(a)^{\alpha}_1,
  ~~k\geq2.
\end{eqnarray}
This can be seen as a fractional generalization of the rising
factorial
\begin{equation}\label{risingfactorial}
    (a)_k=(a+k-1)(a+k-2)...a.
\end{equation}
And the series (\ref{fconfhyperf}) can be seen as a generalization
the confluent hypergeometric function. If $\alpha=1$, it is exactly
the confluent hypergeometric function.

For the fractional Gauss hypergeometric function, consider the
following series
\begin{equation}\label{fhyperf}
    y(z)=\sum_{k=0}^{\infty}\frac{(a)^{\alpha}_k(b)^{\alpha}_k}{(c)^{\alpha}_k}\frac{1}{\Gamma(k\alpha+1)}z^{\alpha\cdot
    k},
\end{equation}
which reduces to the Gauss hypergeometric function when $\alpha=1$.

The ratio of successive coefficients is
\begin{equation}\label{}
    \frac{c_{k+1}\cdot\Gamma(k\alpha+\alpha+1)}{c_{k}\cdot\Gamma(k\alpha+1)} =
    \frac{\bigg(a+\frac{\Gamma(k\alpha+1)}{\Gamma(k\alpha-\alpha+1)}\bigg)\bigg(b+\frac{\Gamma(k\alpha+1)}{\Gamma(k\alpha-\alpha+1)}\bigg)}{\bigg(c+\frac{\Gamma(k\alpha+1)}{\Gamma(k\alpha-\alpha+1)}\bigg)}.
\end{equation}
Making some manipulation, one can get
\begin{eqnarray}
  &&c_{k+1}\cdot
    c\frac{\Gamma(k\alpha+1)}{\Gamma(k\alpha-\alpha+1)}+c_{k+1}\cdot
    \frac{\Gamma(k\alpha+\alpha+1)}{\Gamma(k\alpha-\alpha+1)} \nonumber \\
  &&=c_k\cdot
    ab+c_k\cdot(a+b)\frac{\Gamma(k\alpha+1)}{\Gamma(k\alpha-\alpha+1)}+c_k\cdot\frac{\Gamma(k\alpha+1)}{\Gamma(k\alpha-\alpha+1)}\frac{\Gamma(k\alpha+1)}{\Gamma(k\alpha-\alpha+1)}.
\end{eqnarray}

This equation can be translated to a fractional differential
equation
\begin{eqnarray}
  && ab\cdot f(z)+(a+b)(z-z_0)^{\alpha}D^{\alpha}_{z_0|z}f(z)+(z-z_0)^{\alpha}D^{\alpha}_{z_0|z}\big[(z-z_0)^{\alpha}D^{\alpha}_{z_0|z}f(z)\big] \nonumber \\
  && =c\cdot D^{\alpha}_{z_0|z}f(z)+(z-z_0)^{\alpha}(D^{\alpha}_{z_0|z})^2
  f(z).
\end{eqnarray}
When $\alpha=1$, this equation reduces to the ordinary Gauss
hypergeometric equation. One can check $y(z-z_0)$ defined in
(\ref{fhyperf}) satisfies this equation.

For generalized fractional hypergeometric series
\begin{equation}\label{gfhyperf}
    y(z)=\sum_{k=0}^{\infty}\frac{(a_1)^{\alpha}_k...(a_p)^{\alpha}_k}{(b_1)^{\alpha}_k...(b_q)^{\alpha}_k}\frac{1}{\Gamma(k\alpha+1)}z^{\alpha\cdot k},
\end{equation}
making repeated use of $(z-z_0)^{\alpha}D^{\alpha}_{z_0|z}$, one can
also get its differential equation.

\section{Functions with two fractional indices}\label{FtwoIn}

Not all functions are infinitely fractionally-differentiable, so it
is meaningful to investigate finitely fractionally-differentiable
functions. In Example 1 in Section \ref{FTS}, we have given an
example function that is not infinitely fractionally-differentiable.

We observe that the non-infinitely fractionally-differentiability is
due to another fractional index $\beta$ (no $l$ satisfying
$\beta=l\alpha$). A function $f(x)$ is said to have the behavior of
fractional index $\alpha$ in the right neighborhood of $a$, if it
can be expanded as a series with the basis \{$(x-a)^{m\alpha}|m=
0,1,2,3...$\}.

We also observe that some functions can be expanded into a form with
two fractional indices, i.e.
\begin{equation}\label{twoInSer}
    f(x)=\sum_{m=0}^{\infty}\sum_{n=0}^{\infty}c_{mn}(x-a)^{m\alpha+n\beta}.
\end{equation}
These functions can reduce to a function that is infinitely
$\alpha$-differentiable or infinitely $\beta$-differentiable, but
generally they are finitely fractionally-differentiable functions.

When we are solving fractional differential equations, we should
take care if $[\alpha]>0$ and $\beta\in N$, for the reason of the
relation (\ref{relation1}). However, for $[\alpha]=0$ or $\beta$ not
an integer, the above series is fairly a good ansatz.

A function $f(x)$ is said to be $N\cdot\alpha$
fractionally-differentiable at $a$, if $(D^{\alpha}_{a|x})^m
f(x)|_{x=a}$ is finite for $m\leq N$, but infinite for $m=N+1$. Then
it could be expanded as
\begin{eqnarray}
  f(x) &=& c_{00}+c_{10}(x-a)^{\alpha}++c_{20}(x-a)^{2\alpha}+...+c_{N0}(x-a)^{N\alpha} \nonumber \\
       &+&
       \sum_{N\alpha<m\alpha+n\beta<(N+1)\alpha}c_{mn}(x-a)^{m\alpha+n\beta}+\sum_{m\alpha+n\beta\geq(N+1)\alpha}c_{mn}(x-a)^{m\alpha+n\beta},
\end{eqnarray}
in which the first term in the second line should not be zero.

Now let us see how this kind of series expansion could help to solve
fractional differential equations.

Consider the following equation:
\begin{equation}\label{ExamEQU}
    (x-a)^{\alpha}(D^{\alpha}_{a|x})^2f(x)-D^{\alpha}_{a|x}f(x)=\frac{x-a+(x-a)^{\alpha}}{1-x+a}.
\end{equation}

The right hand side can be expanded as:
\begin{equation}\label{RightH}
    \frac{x-a+(x-a)^{\alpha}}{1-x+a}=\sum_{k=1}^{\infty}(x-a)^k+\sum_{k=0}^{\infty}(x-a)^k(x-a)^{\alpha}.
\end{equation}
It is of index $\alpha$ and index $\beta=1$. It is $1\cdot\alpha$
fractionally-differentiable. So $f(x)$ is $2\cdot\alpha$
fractionally-differentiable. We can write
\begin{eqnarray}
  f(x) &=& c_{00}+c_{10}(x-a)^{\alpha}+c_{20}(x-a)^{2\alpha}  \nonumber \\
       &+& 0+c_{11}(x-a)^{\alpha+1}        \nonumber \\
       &+& c_{02}(x-a)^2 \nonumber \\
       &+&
       \sum_{m\alpha+n>3\alpha,(m,n)\neq(0,2)}c_{mn}(x-a)^{m\alpha+n}.
\end{eqnarray}
Here we assume $0<\alpha<1$.

Or more manageably,
\begin{eqnarray}\label{fExpan}
  f(x) &=& c_{00}+c_{10}(x-a)^{\alpha}  \nonumber\\
       &+& \sum_{n=2}^{\infty}c_{0n}(x-a)^n + \sum_{n=1}^{\infty}c_{1n}(x-a)^{\alpha+n}  \nonumber\\
       &+& \sum_{n=0}^{\infty}c_{2n}(x-a)^{2\alpha+n} +
       \sum_{m\geq3}\sum_{n=0}^{\infty}c_{mn}(x-a)^{m\alpha+n}.
\end{eqnarray}

Substituting (\ref{RightH}) and (\ref{fExpan}) into (\ref{ExamEQU}),
one gets
\begin{eqnarray}
  c_{mn} &=& 0, ~~~~~m\geq3; \nonumber \\
  1/c_{2n} &=& \frac{\Gamma(n+2\alpha+1)}{\Gamma(n+1)}-\frac{\Gamma(n+2\alpha+1)}{\Gamma(n+\alpha+1)}; \nonumber \\
  1/c_{1n} &=& \frac{\Gamma(n+\alpha+1)}{\Gamma(n-\alpha+1)}-\frac{\Gamma(n+\alpha+1)}{\Gamma(n+1)}, ~~~n\geq1; \nonumber \\
  c_{10} &=& 0; \nonumber \\
  c_{0n} &=& 0,~~~n\geq1; ~~~c_{00}=f(a).
\end{eqnarray}

Thus we solved the fractional differential equation (\ref{ExamEQU}).

The above procedure can apply generally to fractional differential
equations. For a generic fractional differential equation
$F[(x-a),y(x-a),D^{\alpha}_{a|x}]=0$, we can solve it by the above
procedure summarized as follows.
\begin{enumerate}
  \item Expand the $(x-a)$ part of the equation and find out the indices of $y(x-a)$,
  \item Find out the number $N$ such that $y(x-a)$ is $N\cdot\alpha$
fractionally-differentiable,
  \item Expand $y(x-a)$,
  \item Substitute the expansion series into the equation,
  \item Find out the coefficients of the expansion series of $y(x-a)$.
\end{enumerate}

We give another example in the following. Consider the fractional
differential equation:
\begin{equation}\label{ExamEQUfg}
    D^{\alpha}_{a|x}y(x)+f(x)y(x)=g(x),
\end{equation}
where $f(x)$ and $g(x)$ are given functions of index $\beta$,
\begin{equation}\label{fcExpan}
    f(x)=f_0+f_1\cdot(x-a)^{\beta}+f_2\cdot(x-a)^{2\beta}+f_3\cdot(x-a)^{3\beta}+...,
\end{equation}
\begin{equation}\label{gcExpan}
    g(x)=g_0+g_1\cdot(x-a)^{\beta}+g_2\cdot(x-a)^{2\beta}+g_3\cdot(x-a)^{3\beta}+...,
\end{equation}
where $f_i$ and $g_i$ ($i=0,1,2,3,...$) are constants.

So $y(x)$ have two indices $\alpha$ and $\beta$. If $g_1\neq0$ and
$\alpha<\beta<2\alpha$, $g(x)$ is $1\cdot\alpha$
fractionally-differentiable, then $y(x)$ is $2\cdot\alpha$
fractionally-differentiable. We can write
\begin{eqnarray}
  y(x) &=& c_{00}+c_{10}(x-a)^{\alpha}+c_{20}(x-a)^{2\alpha}  \nonumber \\
       &+& 0+c_{11}(x-a)^{\alpha+\beta}     \nonumber \\
       &+& c_{02}(x-a)^{2\beta} \nonumber \\
       &+&
       \sum_{m\alpha+n\beta>3\alpha,(m,n)\neq(0,2)}c_{mn}(x-a)^{m\alpha+n\beta}.
\end{eqnarray}

Or more manageably,
\begin{eqnarray}\label{yExpan}
  y(x) &=& c_{00}+c_{10}(x-a)^{\alpha}  \nonumber\\
       &+& \sum_{n=2}^{\infty}c_{0n}(x-a)^{n\beta} + \sum_{n=1}^{\infty}c_{1n}(x-a)^{\alpha+n\beta}  \nonumber\\
       &+& \sum_{n=0}^{\infty}c_{2n}(x-a)^{2\alpha+n\beta} +
       \sum_{m\geq3}\sum_{n=0}^{\infty}c_{mn}(x-a)^{m\alpha+n\beta}.
\end{eqnarray}

Substituting (\ref{fcExpan}), (\ref{gcExpan}) and (\ref{yExpan})
into (\ref{ExamEQUfg}), one gets
\begin{eqnarray}
  c_{00} &=& y(a),~~~~~~~~~~~~~~~~ c_{0n}=0, ~~~~n\geq1; \nonumber  \\
  c_{1n} &=& \frac{\Gamma(n\beta+1)}{\Gamma(\alpha+n\beta+1)}(g_n-f_n c_{00}); \nonumber \\
  c_{2n} &=& -\frac{\Gamma(\alpha+n\beta+1)}{\Gamma(2\alpha+n\beta+1)}(f_n c_{10}+c_{1n}); \nonumber \\
  c_{mn} &=& -\frac{\Gamma((m-1)\alpha+n\beta+1)}{\Gamma(m\alpha+n\beta+1)}c_{(m-1)n} \nonumber \\
         &=& (-1)^{m-2}\frac{\Gamma(2\alpha+n\beta+1)}{\Gamma(m\alpha+n\beta+1)}c_{2n},~~~~~~~~ m\geq3.
\end{eqnarray}

Thus we solved the fractional differential equation
(\ref{ExamEQUfg}).

In this section we have discussed functions with two fractional
indices, but the extension to functions with more fractional indices
will not be difficult.

\section{Summary}\label{Summary}

In summary, in this paper we introduced fractional series expansion
method to fractional calculus. We defined a kind of fractional
Taylor series of infinitely fractionally-differentiable functions.
Based on our definition we generalized hypergeometric functions and
derived their differential equations. For finitely
fractionally-differentiable functions, we observed that the
non-infinitely fractionally-differentiability is due to more than
one fractional indices. We expanded functions with two fractional
indices and illustrated how this kind of series expansion can help
to solve fractional differential equations.



\end{document}